\documentclass[reprint]{revtex4-2}
\usepackage{physics}
\usepackage{graphicx}% Include figure files
\usepackage{dcolumn}% Align table columns on decimal point
\usepackage{bm}% bold math
\usepackage{xcolor}
\usepackage{booktabs}
\usepackage{tabu}
\usepackage{placeins}
\usepackage{hyperref}% add hypertext capabilities
\usepackage[version=3]{mhchem} % Formula subscripts using \ce{}
\begin{document}

\preprint{APS/123-QED}

\title{Efficient Quasiparticle Determination beyond the Diagonal Approximation via Random Compression}

\author{Annabelle Canestraight}
\email{acanestraight@ucsb.edu}
\affiliation{Department of Chemical Engineering, University of California, Santa Barbara, CA 93106-9510, U.S.A.}%

\author{Xiaohe Lei}
\affiliation{Department of Chemistry and Biochemistry, University of California, Santa Barbara, CA 93106-9510, U.S.A.}%

\author{Khaled Z. Ibrahim}%
\affiliation{Applied Mathematics and Computational Research Division, Lawrence Berkeley National Laboratory,
Berkeley, CA 94720, USA}

\author{Vojt\v{e}ch Vl\v{c}ek}%
 %\email{vlcek@ucsb.edu}
\affiliation{Department of Chemistry and Biochemistry, University of California, Santa Barbara, CA 93106-9510, U.S.A.}%
\affiliation{Department of Materials, University of California, Santa Barbara, CA 93106-9510, U.S.A.}%

\date{\today}% It is always \today, today,

\begin{abstract}
Calculations of excited states in Green's function formalism often invoke the diagonal approximation, in which the quasiparticle states are taken from a mean-field calculation. Here, we extend the stochastic approaches applied in the many-body perturbation theory and overcome this limitation for large systems in which we are interested in a small subset of states. We separate the problem into a core subspace, whose coupling to the remainder of the system environment is stochastically sampled. This method is exemplified on computing hole injection energies into CO$_2$ on an extended gold surface with nearly 3000 electrons. We find that in the extended system, the size of the problem can be compressed up to $95\%$ using stochastic sampling. This result provides a way forward for self-consistent stochastic methods and determining Dyson orbitals in large systems.
\end{abstract}

\maketitle

\paragraph*{Introduction}

Single particle states are frequently used in the study of excitation phenomena such as photoionization, electron injection, and generally optical transitions\cite{IvanovLaserCooling,IvanovLaserCooling2,KrylovProbing,PushnigReconstruction,TretiakDensityMat,Luftner2014-rc,martin_2016,OnidaRevModPhys,lei2023exceptional}. The physical interpretation of such single particle states often depends on the specific type of observables\cite{KrylovOrbitals,martin_2016}. In particular, Dyson orbitals, which correspond to the probability amplitude distribution of a specific electron or hole excitation (i.e., quasiparticle state), are directly accessible via orbital tomography and provide insights into the relation between energies and real-space distribution of single particle excitation\cite{PushnigEnergyOrdering,MartinNaturalOrbitals}. This has fundamental implications for chemistry -- e.g., hybridization of quasiparticles on surfaces governs the propensity for direct injection of an electron \cite{lei2023exceptional}. These are just a few compelling reasons to account for the physically meaningful orbital distrinutions, especially for problems concerning (chemical) interfaces.

In practice, however, single-particle states for interfacial systems are typically taken from the Density Functional Theory (DFT) \cite{wu2021identification,biller2011electronic,egger2015reliable}, as the cost of higher level theory is too high. While DFT can handle extremely large systems\cite{Millionelectrons}, these calculations can not, even in principle, yield quasiparticle (QP) energies or the Dyson orbitals\cite{martin_2016,Engel&Dreizler}. A natural and widely applied extension, especially in condensed matter problems, is application of the Many Body Perturbation Theory (MBPT) employing Green's function formalism\cite{hedin1965new,martin_2016,Hybertsen&LouieOG,ShamGW}. In particular, the $GW$ approximation, which truncates the correlation expansion to non-local charge density fluctuations, has emerged as arguably the most popular approach\cite{OnidaReview,GolzeCompendium} and higher order corrections emerged recently\cite{CarlosMultiQP,CarlosGWGamma,StochasticVertexcorrections,Gwen_embeddingvertex}. Its self-consistent solution yields both QP energies and the Dyson orbitals \cite{rusakov2016self,caruso2013self,Kaplan2015}.  However, it is common to apply $GW$ approach as a one-shot correction,  $G_0W_0$, employing the Kohn-Sham Green's function $G_0$ and the screened coulomb interaction $W_0$ derived from the underlying Kohn Sham DFT solutions. Despite its approximate nature, $G_0W_0$ often provides good estimates of band gaps\cite{YangHedinShift,sGWMolecules,Kaplan2015,rignanese2001quasiparticle,Rohlfing1993MoreMaterials,MartinMols}.  The use of one-shot corrections has been largely motivated by the computational cost, which scales as  $\mathcal{O}(N^4)$  with the number of electrons in conventional implementations\cite{pham2013g,Galli2015}. The computational cost has been significantly decreased by stochastic sampling approaches in $GW$ (and post-$GW$) to be nearly linear; 1000's of states can thus be studied\cite{vlvcek2018swift,VojtechStochGW,GwenDonor,romanova2020decomposition,weng2021efficient}. 
However, even in the stochastic $GW$, ``updating'' the single-particle basis (i.e., finding the Dyson orbitals) is difficult\cite{Romanova2022} and, in practice, usually avoided\cite{Vlcek2018}. Routine calculations of QP orbitals in realistic systems with thousands of electrons are still elusive. This is true even if one is, in principle, interested in treating a \textit{small subset} of states, as exemplified in this work (see below).

\begin{figure}[ht]
    \includegraphics[width=.8\linewidth]{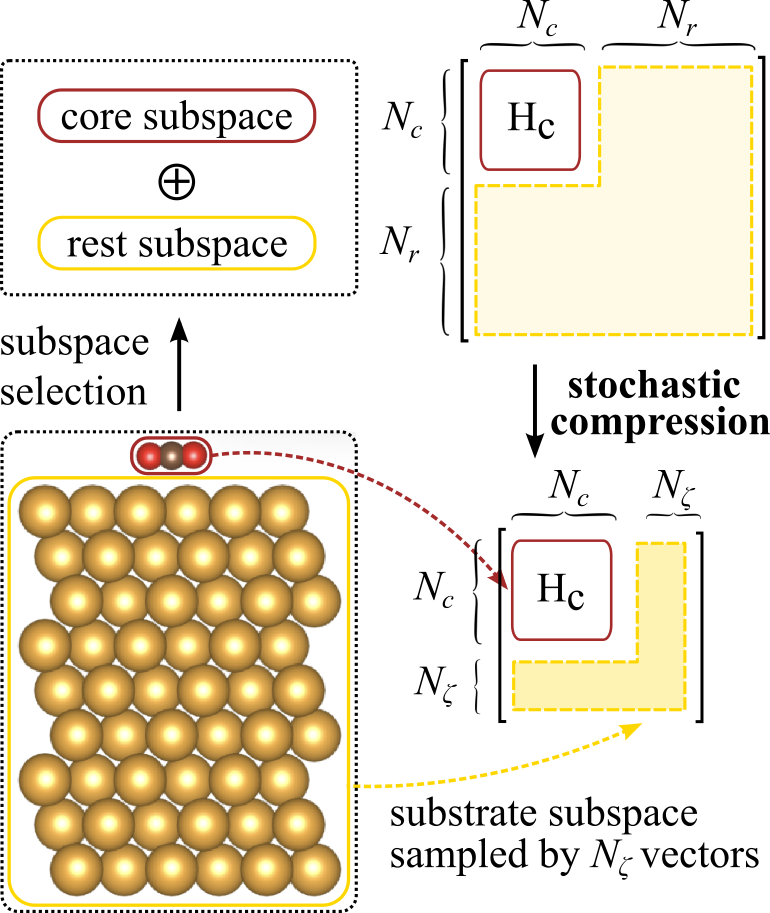}
    \caption{Illustration of the stochastic compression technique, which samples the ``rest subspace'' using a set of (filtered) random vectors, here spanning the single particle states of the gold substrate. }
    \label{fig:fig1}
\end{figure}

Here, we tackle this problem and present a scheme without the diagonal approximation for realistic nanoscale systems. This stochastic framework is exemplified for CO$_2$ molecule on a large Au slab. For this problem, the surface contributions to the orbitals are sampled, drastically reducing the cost of QP calculations. This method divides the system into a set of states in an ``core'' subspace, treated by standard stochastic MBPT, and a rest space, for which additional sampling is introduced. This step is combined with a search over the fixed-point solutions of the frequency-dependent QP Hamiltonian, which is basis representation independent and thus enables the use of random vectors.

We apply these methods to a prototypical system of a small molecule on a plasmonic surface (\ce{CO_2} on Au illustrated in the inset in Fig.~\ref{fig:fig1}). In the practical demonstration for an extended Au (111) surface with 270 atoms (2986 electrons), we found convergence in the hybridized HOMO energy with a $95\%$ rank compression compared to evaluation in the full canonical orbital basis. This success provides a way to use costly high-level theories to study realistic chemical systems.

\paragraph*{Formalism}

The time-ordered Green's function (GF) contains information about the quasiparticle (QP) energy spectrum and lifetimes, and it corresponds to the probability amplitude of a QP propagation between two space-time points $\mathbf{r},t$ and $\mathbf{r}',t'$.  In the Lehmann representation, it is expressed as
\begin{equation}
\begin{aligned}
    {G}(\mathbf{r},\mathbf{r'},\omega) = \sum_n  \bigg[ \frac{\psi_n (\mathbf{r}) \psi_n(\mathbf{r'})^*}{\omega-\varepsilon_n-i\eta}
 \bigg]   ,
\end{aligned}
\end{equation}
where the Dyson orbitals are obtained as $\psi_n (\mathbf{r}) = \left\langle \Psi^n_{N-1}\middle | \hat \psi({\mathbf r}) \middle | \Psi_N^0\right\rangle$ from the $N$-particle ground state and the $n^{\rm th}$ excited state of the $N-1$ particle system, where $ \hat \psi({\mathbf r})$ is the field operator. The poles of the GF are located at the QP energies,  $\varepsilon_n$, here corresponding to the charge removal \cite{martin_2016}. Charge addition is treated analogously.  
 The GF poles are conveniently expressed as solutions to a non-linear eigenvalue problem for an effective Hamiltonian obtained by downfolding interactions with the system\cite{martin_2016}:
\begin{equation}
    \hat{H}_{QP}(\omega)\left| \psi\right\rangle=\omega \left| \psi \right\rangle
    \label{nlev}
\end{equation} 

In practice, the QP Hamiltonian is divided into a static and local term, $H_0$, which typically contains all one-body contributions, while a space-time non-local portion is represented by the self-energy operator $\Tilde{\Sigma}$\cite{GolzeCompendium}. The latter is approximated by selected types of interaction diagrams (and their resummation). As $\tilde \Sigma$ is conceptually equivalent to the exchange-correlation potential applied in the Kohn-Sham density functional theory (KS DFT), the QP Hamiltonian is practically constructed as a perturbative correction on top of such a mean-field starting point:
\begin{equation}\label{HQP}
    \hat{H}_{QP}(\omega)=\hat{H}_{0,\rm KS} -\hat{V}_{xc}+\hat{\Sigma}(\omega),
\end{equation}
where $\hat{H}_{0,\rm KS}$ is the KS DFT Hamiltonian.

Further, the ``one-shot'' correction corresponds to:
\begin{equation}
    \Sigma(\mathbf{r},\mathbf{r'},\omega)= i\int \frac{d\omega'}{2\pi}{G}_0(\mathbf{r},\mathbf{r'},\omega+\omega') W_0 (\mathbf{r},\mathbf{r'},\omega') ,
\end{equation}
where $\mathcal{G}_0$ has poles at the DFT Kohn Sham eigenvalues, $\varepsilon_0$, and $W_0$ is the screened coulomb interaction. The self-consistency requires repeated construction of $\Sigma$ and re-evaluation of Eq.~\ref{nlev}; multiple flavors of self-consistent approaches have been developed \cite{rusakov2016self,caruso2013self}. Typically, the convergence pattern is smooth. If the KS DFT single-particle states are close to the Dyson orbitals, the ``one-shot'' correction provides good estimates of QP energies, yet the quality of the mean-filed eigenstates are not \textit{a priori} known. 

A step beyond this practice is to diagonalize  $H_{QP}$ in Eq.~\ref{nlev} in the orbital basis, yielding Dyson orbitals (in the first iteration) and updated one-shot QP energies in the $GW$ approximation\cite{martin_2016}. Note that, in principle, the nonlinear problem in Eq.~\ref{nlev} holds for multiple values of $\omega$ associated with satellite features \cite{CarlosMultiQP,satellites1,satellites2}. In this work, we will focus only on the primary QP peaks, i.e., we seek a single solution to the QP Hamiltonian in the vicinity of $\varepsilon_0$ and look for the fixed point solutions to $    \omega_i=\bra{\phi_i}\hat{H}_{QP}[\omega_i]\ket{\phi_i}$. Note that $H_{QP}$ is non-hermitian, and each QP state, in general, corresponds to $H_{QP}$ computed at a different frequency.
 
In practical schemes\cite{Kaplan2015,MVSandKotani,SCGW2,Vlcek2018}, it is common to construct a single ``static'' effective Hamiltonian (yielding orthogonal eigenstates). However, due to the non-linearity of this problem, it is not entirely clear at what frequency the self-energy should be evaluated. For strongly diagonally dominant $H_{QP}$, i.e., those where KS DFT orbitals are, in fact, close to the Dyson orbitals, one may evaluate $\omega_i$ as the fixed point solution for the diagonal entries. The remaining off-diagonal self-energy is e.g., $\Sigma_{ij}=\frac{1}{4}\left[\Sigma_{ij}(\omega_i)+ \Sigma_{ji}(\omega_i)+\Sigma_{ij}(\omega_j)+\Sigma_{ji}(\omega_j)\right]$. In this form, it is possible to construct a static and hermitized QP Hamiltonian. By enforcing the hermicity of $H_{QP}$, we impose that the resulting QP states are orthonormal. The QP energies are purely real, corresponding to an infinite lifetime QP. Alternatively, one can therefore relax the latter step by taking $\Sigma_{ij}=\frac{1}{2}\left[\Sigma_{ij}(\omega_i)+ \Sigma_{ij}(\omega_j)\right]$. 

Note that both approaches strongly depend on the basis choice. We illustrate this in detail in the Supporting Information (SI) for the acrolein molecule, for which the magnitudes of the off-diagonal terms are $98.5\%$ smaller than the diagonal ones for the canonical KS DFT basis. The situation changes dramatically when localized (unitary transformed) orbitals are employed. Hence, depending on the construction of a single $H_{QP}$, the resulting QP energies change by as much as $10\%$ and translate to changes of 0.77~eV on average for acrolein. 

Since our goal is to determine Dyson orbitals for a selected subspace of interest (which will be constructed from localized basis states), we avoid any approximation to the fixed point solution. In this method, the whole QP Hamiltonian is evaluated at multiple frequencies, and the QP eigenvalues are found as the fixed point solutions to Eq.~\ref{nlev}. No assumptions are further made about the hermicity of the Hamiltonian matrix; a graphical example of such a fixed point solution for the $H_{QP}$ is also illustrated in the SI.

\paragraph*{Stochastic Compression of QP states}

When studying a large system with a subspace of particular interest, it is prohibitively expensive to employ all $M$ electronic states. It is also insufficient to assume that the Hamiltonian matrix takes a block-diagonal form due to the coupling between the subspace and its orthogonal complement. To handle such a case, we propose a method of stochastic matrix compression where a portion of the 
$H_{QP}$ matrix is represented by a set of random vectors. These vectors sample a large portion of the Hilbert space, which overall contributes to the QP shift and affects the Dyson orbitals, but for which each individual single particle state has only a limited contribution. 

As illustrated in Fig. \ref{fig:fig1}, we separate the ``core subspace'' spanned by $N_c$ deterministic states, $\{\phi^c\}$,  (e.g., the original KS DFT eigenstates), and the remainder spanned by a $N_s$ stochastic states $\{\zeta\}$, constructed as a random linear combination of the KS states that are orthogonal to the $\{\phi^c\}$ set: $ \ket{\zeta}=\sum_{i=1\bot \phi^c}^{N_s}c_i \ket{\phi_i}$.
In the final step, the individual random states are orthogonalized via the Gram-Schmidt process. Because this change of basis is guaranteed to be a unitary transformation of the Hamiltonian matrix, when the whole system is diagonalized, the resulting eigenstates will be the same. 
When the Hamiltonian matrix is truncated in this new stochastic basis, the coupling of each stochastic state to the core subspace will represent the subspace interaction with the full environment. In this way, we have ``compressed'' the information of the whole system environment into a single state. Given that the fixed point solution is basis independent (as illustrated in Fig. S.3), a total number of states $N_c + N_s$ is the same as the dimension of $H_{QP}$, $M$, we necessarily obtain the same QP energies. For fewer random states, $N_c+ N_s< M$, and the computation is less expensive. Note that the QP energy has a finite statistical error, which decreases as $1/\sqrt{N_s}$ with the number of states sampling the off-diagonal self-energy contributions. As we show below, the convergence of the QP energies is smooth. Further, note that instead of the canonical single particle states in the above equation, we achieve further speedup if already preselected (filtered) subset of states (orthogonal to the $\{\phi^c\}$) are used in the construction of $\ket{\zeta}$.

\paragraph*{Results}
We now demonstrate the method practically for the CO$_2$ molecule on the Au substrate for which we intend to extract the energies of quasi-hole states on the molecule (i.e., corresponding to the charge removal energies from CO$_2$ on the surface). We first construct a minimal example on which we can solve entirely and illustrate how stochastic sampling smoothes the convergence of the QP energies. Later, we show a realistic example with nearly 3,000 electrons, which cannot be easily solved without the sampling methodology.

%Having developed a way to diagonalize the Hamiltonian in an arbitrary basis, we will now demonstrate its advantage in a solvable model system. We investigate how a change of basis can allow us to divide the system and use few states in the diagonalization. 

We will demonstrate the success of our stochastic sampling method on a minimal system of \ce{CO_2} on a bilayer of 8 gold atoms. This system contains only 52 occupied states, which we also treat explicitly. Note that, in principle, the hybridization extends beyond merely the occupied manifold, but to illustrate the methodology, we consider only the rotation within the occupied subspace. To see the surface-induced changes, calculate the QP states for a \ce{CO_2} molecule in a vacuum ($N=8$) and for the minimal composite system ($N=52$). We find that the seven lowest valence states of the molecule shift in energy when the substrate is included, but the eigenvectors (orbitals) do not change in response to the gold substrate. %\vv{These seven states are separated by more than XXeV from the Fermi level and thus remain purely molcule-like}. 

In contrast, HOMO state behaves differently: no single state would correspond to the molecular HOMO (either the canonical DFT or Dyson orbitals computed at the $G_0W_0$ level). Instead, there are multiple \textit{hybridized} states sufficiently localized on the molecule, whose eigenvalues lay within a small range of energies. We aim to characterize them and, consequently, to find a characteristic QP energy for this distribution of HOMO QP for the \ce{CO_2} molecule on Au. 

We thus define a ``core subspace'' comprising the states with the most molecular character. In practice, they are identified based on projection onto localized (unitary transformed) orbitals centered on \ce{CO_2}, e.g., using the molecular reconstruction technique which is applied here\cite{Gwenreducedscaling,weng2021efficient}. The corresponding projection value is:
\begin{equation}\label{eq:projector}
   P_i=\sum_j|\braket{\xi_j}{\phi_i}|^2
\end{equation}
Here, $\{\ket{\xi}\}$ and {$\{\ket{\phi}\}$} are the sets of transformed (localized) and canonical KS DFT states respectively. Each KS state with $P$ greater than a chosen threshold is included in the core region. This preselection separates the ``core'' subspace from the rest.

\begin{figure}
    \centering
    
    \includegraphics[width=\linewidth]{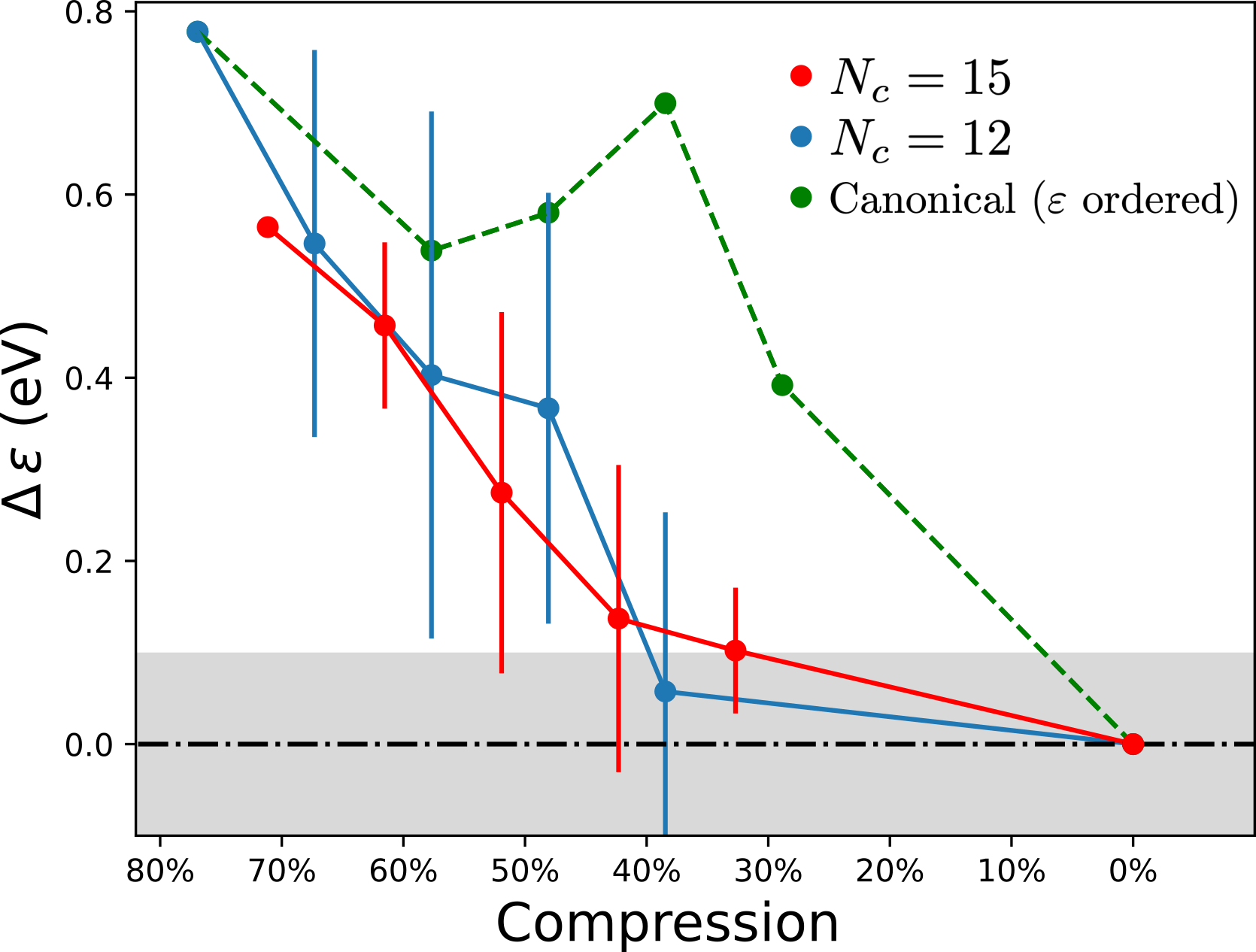}

    \caption{\textbf{Hybridized HOMO Convergence (Minimal System)}: Core sizes of $N_c=12$ and $N_c=15$ are used, with the remaining states sampled with equal weight. In contrast, the adding the states by energy ($\varepsilon$ ordered)  demonstrates the lack of smooth convergence. The gray-shaded region shows where the spectrum converges within 0.1 eV}
    \label{fig:minimalplt}
\end{figure}

We now track the fixed point HOMO QP solution with the number of states considered in the $H_{QP}$, i.e., we gradually add more states outside of the core subspace. The molecular HOMO is hybridized with many of the surface states. We thus define a single energy for this state by taking its mean value, constructed by weighting by the projection onto the HOMO of \ce{CO_2} in a vacuum.
The results are shown by the green color in Fig.~\ref{fig:minimalplt}. The left-most point represents only the core space, containing 12 orbitals corresponding to 23\% of the entire $H_{QP}$. The size of the problem is increased by adding states depending on their distance from the KS DFT HOMO energy, as one would expect that the hybridization of states will be small for energetically distant states. This does not produce a smooth convergence (green line in Fig.~\ref{fig:minimalplt}) as some surface hybridization is due to Au orbitals that are far from the core subspace.

%We believe that tackling the problem of system divisibility on a highly hybridized system represents an ultimate challenge for the method.

To demonstrate the stochastic approach, we now instead sample the remaining KS states using random vectors: 
\begin{equation}\label{eq:zeta_rand_mix}
    \ket{\zeta}=\frac{1}{\sqrt{N_e}}\sum_{j=1}^{N_e}w_j e^{i\theta_j}\ket{\phi_j}
\end{equation}
Here $\theta_j \in [0,2\pi]$ is randomly chosen, and $N_e$ is the number of ``rest'' states used with weight $w_i$. Note that we can either sample all remaining states evenly ($w_i = 1 \forall i$), but generally, we consider a random selection from a distribution within the sampled subspace (determined by $P_i$ in Eq.~\ref{eq:projector}) as we show later.

%The Hamiltonian matrix is diagonalized using the full frequency method, and the resulting eigenvectors are evaluated to determine which are part of the hybridized HOMO spectrum. We do so by projecting the eigenvectors onto the molecular wannier states again. The average of this energy spectrum is taken and presented in \ref{fig:minimalplt}.

Once we have obtained the set $\{\zeta \}$, we randomly draw $N_s$ of them and the fixed point solution is then found for $H_{QP}$ with the dimension of $N_c + N_s$.  The results for $N_c =12$ and variable $N_s$ is in Fig.~\ref{fig:minimalplt} and shows a monotonic and smooth convergence towards to the asymptotic value (obtained for the entire 52 occupied states). The stochastic sampling was repeated ten times for each step with a different set of $N_s$ random vectors; the standard deviations are indicated in the plot and they naturally disappear in the complete basis limit. For instance, for $N_s =20$, i.e.,  62\% of the entire system, we see a difference of $0.057\pm 0.19$ eV between the mean HOMO QP energies. For an increased core space, $N_c =15$, we see that the HOMO QP value converges similarly, i.e., the size of the core space is not changing the convergence profile significantly. For $N_s=20$ (i.e., 32.5$\%$ compression of matrix rank), the resulting spectrum mean falls within a 100~meV from the value obtained from the diagonalization of the full matrix. 

%he size reduction is limited as there is little redundancy  

%Typically, we estimate the QP energies with accuracy of 50~meV which is achieved for $N_s\approx$ if we linearly interpolate result with the final solution, we would predict that X stochastic states are required to achieve the desired accuracy of $0.05$ eV. 

% This target accuracy of $.05$ eV is defined arbitrarily as a goal for the method of stochastic compression. We wish to remind the reader, however, that the stochastic error of the $GW$ calculations is $.15$ eV, implying that the method described here surpasses the overall method in precision. Nevertheless, our method has been developed to work independently of how the GW calculation is performed.

% To further demonstrate the success of the stochastic matrix compression, we repeat the procedure by including the remaining stochastic states in order of increasing energy, rather than by drawing stochastic states. We justify this comparison by our naive assumption that the states with any molecular character will have energies near to the core state energies. This curve does not decrease monotonically in distance from the true spectral mean. This division of states provides no means for reducing the overall size of the Hamiltonian matrix. 

Without any prior knowledge or arbitrary truncation of the KS states, we can capture molecule-surface hybridization effects by employing stochastic states representing the substrate environment. This description is systematically improvable by increasing both $N_c$ and $N_s$. In general, the cost reduction provided by the stochastic sampling is due to circumventing the summation over many states that contribute either similarly or very little to the expectation values in question\cite{StochasticReview}. For a small system such as the one used here, the amount of compression is less significant as most of the states contribute to the QP HOMO energy. 

%As the threshold for the projection $P$ is lowered  core subspace is lowered, we can systematically improve the result, relying less heavily on the stochastic states. While the system studied here is small and can be computed in full, it provides a demonstration the computational cost can be reduced when the method is applied to a large and realistic system. 

%\subsection{Extended System}

We now turn to a realistic large-scale system for which such a calculation would not be possible with standard methods. Here, we study \ce{CO_2} molecule on an extended Au-111 surface of 270 atoms, containing 2986 electrons. The system is treated analogously to the minimal example: we selected a core subspace of 15 and 25 states. Due to the molecule-surface hybridization, $N_c=15$ is the minimal size of the core space identified for this particular problem. Next, the stochastic sampling uses a filtered distribution in Eq.~\ref{eq:zeta_rand_mix} in which we consider a linear combination of states that are sufficiently localized on the molecules. In practice, this step determines the sampled subspace, which is practically restricted to states with  $P$ greater than  a selected threshold, $P_T$. Here we consider two cases $P_T= 10^{-3}$ and $P_T= 5\times10^{-4}$.

\begin{figure}
    \centering
    \includegraphics[width=\linewidth]{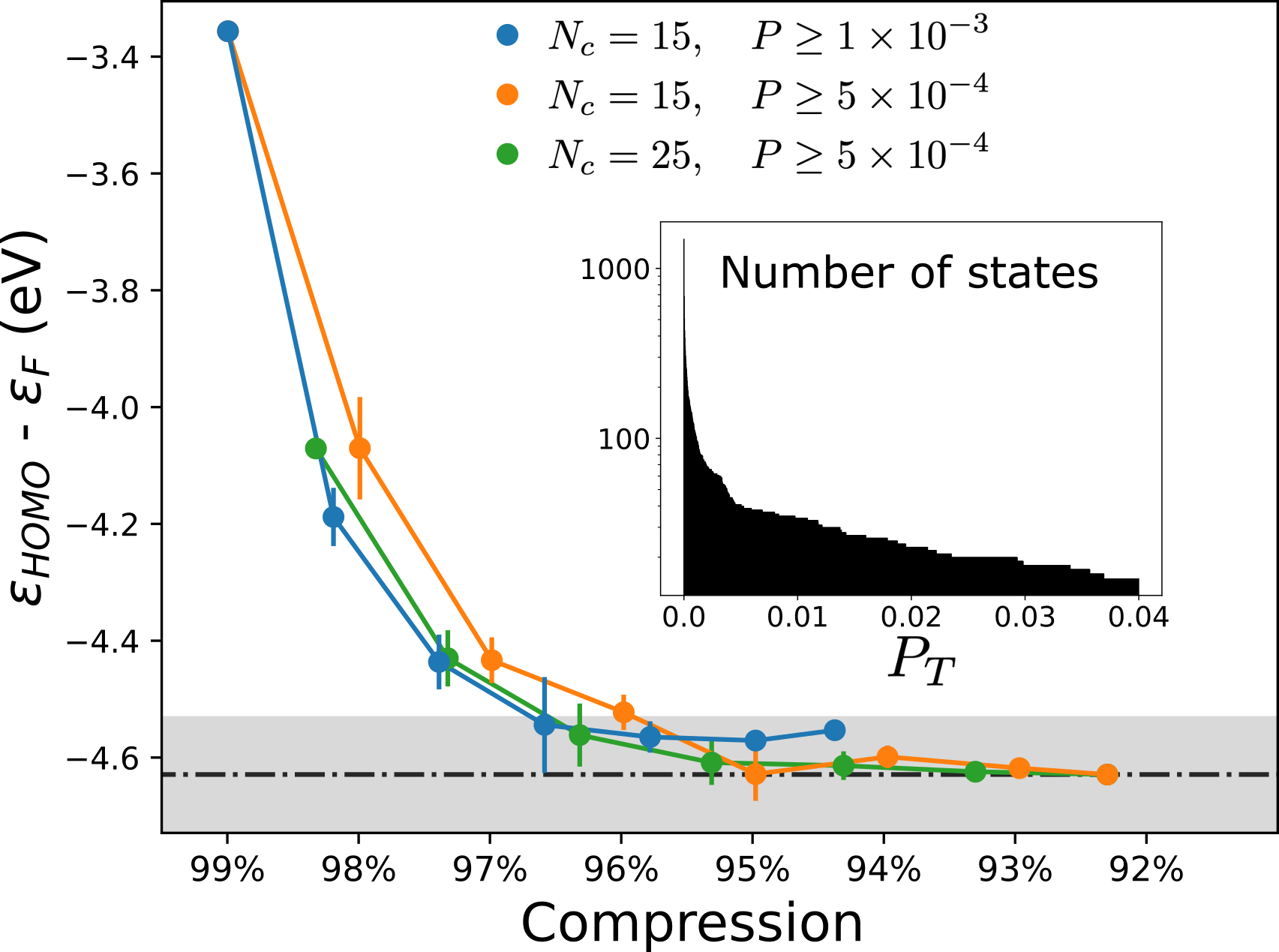}
    \caption{\textbf{Hybridized HOMO Convergence (Large System with 2986 electrons):} Convergence is demonstrated across multiple core sizes and stochastic sampling filterings. The shaded region indicates that the HOMO is within 0.1 eV of the converged value. The inset shows the log-scale histogram of the number of states for which the projection value $P \geq P_T$, where $P_T$ is the filtering cutoff (see text).}
    \label{extendedplts}
\end{figure}

From Fig.~\ref{extendedplts} we can see that the HOMO energy converges with only 5$\%$ of the total number of states used\footnote{In practical execution on HPC machines, namely at the National Energy Research Scientific Computing Center using a Intel Xeon Processor E5-2698 v3 at 2.3 GHz, the difference in computational cost for using $96\%$ and $92\%$ compressed matrix amounts to $\sim$77,000 CPU$\cdot$hrs. For the maximally compressed converged calculation (96\% compression), the entire calculation amounts to $\sim$ 60,000 CPU$\cdot$hrs.}. For slightly increased selectivity (i.e., lower projection threshold $P$), the stochastic sampling of the hybridization converges similarly. Further, the size of the core subspace does not significantly impact the convergence rate: when $N_c=25$ with the filtering threshold of $P_T =5\times10^{-4}$, the curve matches that of the $N_c=15$ for the same value of $P_T$. This suggests that the size of the core subspace can be decreased, possibly at the expense of using more stochastic samplings.

Finally, note that when the orbital re-hybridization is used at the $G_0W_0$ level, the HOMO QP energy moves down in energy by more than 1~eV. Since approximate semilocal KS DFT is known to suffer from overdelocalization, it is expected that the physical Dyson orbitals are more localized than the canonical KS DFT eigenstates. In turn, stronger localization of HOMO is typically associated with its energy decrease\cite{lei2023exceptional}. These observations are thus in line with what the MBPT should accomplish and underline the need for more appropriate treatment of surface phenomena.

%of using more  and  that we have maximized our efficiency in solving this problem, and that by selecting the most molecular stochastic states, we accomplish nearly the same thing as we do in expanding the core.

\paragraph*{Outlook} The rapid convergence of the QP energies with $N_s$ implies that when we stochastically sample the matrix, aided by preselection and filtering, we can represent the full QP spectrum for a molecule that hybridizes with an extended surface using less than 5$\%$ of the system. The $H_{QP}$ matrix size is thus compressed by 95$\%$. This is largely due to the significant ``redundancy'' of information encoded in individual single-particle states, and the sampling allows sampling all (or a large filtered portion of them) simultaneously through random vectors. The approach presented here will enable the treatment of large-scale interfacial problems and opens the door for efficient self-consistent stochastic MBPT. 

% \begin{itemize}
%     \item The degree of compression that is possible is dependent on the amount of ``data redundancy" in the system. Even then, we were able to compress by up to 33 percent
%     \item this method is promising for the study of interfaces between large surfaces and small molecules
%     \item the method hinges completely on identifying a subspace based on locality $\rightarrow$ limits us to study localized/molecular states
%     \item further, we saw that in the case of the minimal system we were able to sample the entire set of states and see convergence. In the case of the extended system, we sampled from only states that met a certain locality threshold both before and after the orthogonalization procedure. This was needed because ``important states" were diluted too much to count when the full space was sampled. This begs the question of if it is possible to simply choose a localized basis independent of the stochastic procedure (i probably dont want to say this, but I think it)
%     \item probably forget most of this, but the decrease in energy of the HOMO shows that the dyson orbital has had the delocalization error of the DFT corrected
% \end{itemize}
%

%

\section*{Acknowledgements}
The development (A.C., X.L., and V.V.) of the stochastic compression technique was supported by the NSF CAREER award (DMR-1945098). The implementation and numerical testing (A.C., K.I., and V.V.) is based upon work supported by the U.S. Department of Energy, Office of Science, Office of Advanced Scientific Computing Research and Office of Basic Energy Sciences, Scientific Discovery through Advanced Computing (SciDAC) program under Award Number DE-SC0022198.  This research used resources of the National Energy Research Scientific Computing Center, a DOE Office of Science User Facility supported by the Office of Science of the U.S. Department of Energy under Contract No. DE-AC02-05CH11231 using NERSC award BES-ERCAP0020089.

\medskip

\bibliographystyle{unsrt}
\bibliography{bib}
\end{document}